\documentclass[12pt]{article}
\usepackage{amsfonts}
\usepackage{amsmath}
\usepackage{amssymb}
\usepackage{graphicx}
\usepackage{color}
\usepackage[all, knot]{xy}
\usepackage{tikz}
\usepackage{array}
\usepackage{hyperref}

\usepackage[utf8]{inputenc}
\usepackage{epstopdf}
\usepackage[footnotesize]{caption}
\usepackage{amsthm}
\usepackage{enumitem}
\usepackage{mathrsfs}
\usepackage{mathtools}

\usepackage[margin=3cm]{geometry}

\def \be {\begin{equation}}
\def \ee {\end{equation}}
\def \bea {\begin{eqnarray}}
\def \eea {\end{eqnarray}}
\def \nn {\nonumber}

\def \rr {\raise.35ex\hbox{\small $\prime$}\kern-.17em{\mbox{\large $\imath$}}}

\def \dels {\partial\kern-.6em /\kern.1em}
\def \As {{A\kern-.5em / \kern.5em}}
\def \Ds {D\kern-.7em / \kern.5em}

\def \ks {k\kern-.5em /}
\def \ls {l\kern-.5em /}

\newcommand{\ci}[1]{}

\newcommand{\ba}{\begin{eqnarray}}
\newcommand{\ea}{\end{eqnarray}}
\newcommand{\bal}{\begin{align}}
\newcommand{\eal}{\end{align}}
\newcommand{\bay}[1]{\left(\begin{array}{#1}}
\newcommand{\eay}{\end{array}\right)}

\newcommand{\hide}[1]{}

\newcommand{\fsl}[1]{\ensuremath{\mathrlap{\!\not{\phantom{#1}}}#1}}

\newlist{axioms}{enumerate}{2}
\setlist[axioms,1]{label=\textbf{A\arabic{axiomsi}.}, ref=A\arabic{axiomsi}}
\setlist[axioms,2]{label=\textbf{A\arabic{axiomsi}\rlap{\myEnumCounter{axiomsii}}.},%
                   ref=A\arabic{axiomsi}\myEnumCounter{axiomsii},%
                   align=parleft,%
                   leftmargin=0em,%
                   itemsep=1.4ex,%
                   before={\stepcounter{axiomsi}}}

  \usetikzlibrary{decorations.markings}

\begin{document}
\begin{titlepage}

\begin{center}

\textbf{\LARGE
Non-Hermitian Lattice Fermions in\\ 
2D GNY Model
\vskip.3cm
}
\vskip .5in
{\large
Xingyu Guo$^{a, b,c}$ \footnote{e-mail address: guoxy@m.scnu.edu.cn}, 
Chen-Te Ma$^{b, c, d, e, f, g}$ \footnote{e-mail address: yefgst@gmail.com}, 
and Hui Zhang$^{a, b, c, h}$ \footnote{e-mail address: Mr.zhanghui@m.scnu.edu.cn}
\\
\vskip 1mm
}
{\sl
$^a$ 
Key Laboratory of Atomic and Subatomic Structure and Quantum Control (MOE), Guangdong Basic Research Center of Excellence for Structure and Fundamental Interactions of Matter, Institute of Quantum Matter, South China Normal University, Guangzhou 510006, China. 
\\
$^b$
Guangdong Provincial Key Laboratory of Nuclear Science,\\
 Institute of Quantum Matter,
South China Normal University, Guangzhou 510006, Guangdong, China.
\\
$^c$
Guangdong-Hong Kong Joint Laboratory of Quantum Matter,\\
 Southern Nuclear Science Computing Center, 
South China Normal University, Guangzhou 510006, Guangdong, China.
\\
$^d$
Department of Physics and Astronomy, Iowa State University, Ames, Iowa 50011, US.
\\
$^e$
Asia Pacific Center for Theoretical Physics,\\
Pohang University of Science and Technology, 
Pohang 37673, Gyeongsangbuk-do, South Korea. 
\\
$^f$
School of Physics and Telecommunication Engineering,\\ 
South China Normal University, Guangzhou 510006, Guangdong, China.
\\
$^g$
The Laboratory for Quantum Gravity and Strings,\\ 
Department of Mathematics and Applied Mathematics, 
University of Cape Town, Private Bag, Rondebosch 7700, South Africa.
\\
$^h$
Physics Department and Center for Exploration of Energy and Matter,\\
Indiana University, Bloomington, Indiana 47408, US.
}
\\
\vskip 1mm
\vspace{40pt}
\end{center}
\newpage
\begin{abstract}
We work the lattice fermions and non-Hermitian formulation in the 2D GNY model and demonstrate the numerical implementation for two flavors by the Hybrid Monte Carlo.
Our approach has a notable advantage in dealing with chiral symmetry on a lattice by avoiding the Nielsen-Ninomiya theorem, due to the non-symmetrized finite-difference operator.
We restore the hypercubic symmetry by averaging over all possible orientations with the proper continuum limit.
Our study is the first simulation for the interacting fermion formulated in a non-hermitian way.
We compare the numerical solution with the one-loop resummation.
The resummation results matches with the numerical solution in $\langle\phi\rangle$, $\langle\phi^2\rangle$, $\langle\mathrm{Tr}(\bar{\psi}_1\psi_1+\bar{\psi}_2\psi_2)/2\rangle$, and $\langle\mathrm{Tr}(\bar{\psi}_1\psi_1+\bar{\psi}_2\psi_2)\phi/2\rangle$.
We also used the one-loop resummation to provide the RG flow and asymptotic safety in the 2D GNY model.
\end{abstract}
\end{titlepage}

\section{Introduction}
\label{sec:1}
\noindent 
Gross-Neveu (GN) model describes the Dirac fermion fields interacting via four-fermion interactions in 2D spacetime.
The Lagrangian formulation for one flavor in the 2D Euclidean spacetime is
\bea
S_{\mathrm{GN}}\lbrack\bar{\psi}, \psi\rbrack
=
\int d^2x\ \bigg(\bar{\psi}(\fsl{\partial}+m_{\mathrm{F}})\psi-\frac{g^2}{2}(\bar{\psi}\psi)^2\bigg),
\eea
where $m_{\mathrm{F}}$ is the Dirac fermion mass, and $g^2$ is a dimensinless positive conpuling constant.
When considering $N$ flavors, GN theory is asymptotically free in the large $N$ limit.
This model has a similar property to Quantum Chromodynamics (QCD).
When $N=1$, the model is equivalent to the massive Thirring model \cite{Thirring:1958in}.
The partition function of the massive Thirring model is equivalent to the sine-Gordon model \cite{Skyrme:1958vn,Skyrme:1961vr,Coleman:1974bu,Mandelstam:1975hb,Buscher:1987sk,Buscher:1987qj,Burgess:1993np}.
The sine-Gordon model has a significant ultraviolet (UV) fixed point that results in a finite coupling constant. 
As a result, the GN model for one flavor is {\it asymptotically safe} \cite{Kovacs:2014fwa}. 
Asymptotic safety implies that the theory remains well-defined and finite at arbitrarily high energies or short distances. 
However, the GN model is UV finite, but it cannot be extended to four dimensions. 
Therefore, we are studying the Gross-Neveu-Yukawa (GNY) model in a two-dimensional spacetime. 
The GNY model is described by the action below: 
\bea
S_{\mathrm{GNY}}\lbrack\bar{\psi}, \psi, \phi\rbrack
=
\int d^2x\ \bigg(\bar{\psi}(\fsl{\partial}+m_{\mathrm{F}}+\phi)\psi-\phi\Box\phi+\frac{1}{2g^2}\phi^2\bigg).
\eea 
In the GNY model's action, the term $\phi\Box\phi$ represents the kinetic term for the scalar field, and $m_{\mathrm{B}}^2/2=\phi^2/(2g^2)$ represents its self-interaction term. 
The coupling constant $g^2$ here is also positive, similar to the GN model. 
The GNY model indeed introduces an additional scalar field $\phi$ to the GN model, primarily to ensure {\it renormalizability} in four dimensions. 
The kinetic term associated with the scalar field allows for more control over the model's behavior, particularly in the UV regime. 
While the kinetic term for $\phi$ may influence the UV behavior of the model, the expectation is that the essential property of asymptotic safety, which is characteristic of the GN model for one flavor, will still hold in the GNY model. 
\\

\noindent 
Non-perturbative Quantum Field Theory (QFT) computation in continuous spacetime involves evaluating an infinite-dimensional path integral.
Because the path integration is computationally intractable, we work on a discrete spacetime and consider finite-dimensional path integral \cite{Wilson:1974sk}.
We can compute the path integral by numerical techniques, like the Monte Carlo (MC) method.
The lattice formulation is a practical simulation method used to study non-perturbative physics. 
The continuum field theory is recovered in the limit of infinite lattice size and infinitesimal spacing (a).
{\it Nielsen-Ninomiya theorem} \cite{Nielsen:1980rz, Nielsen:1981xu, Karsten:1981gd} poses constraints on constructing a {\it hermitian} $d$-dimensional lattice action for fermions $S_{\mathrm{F}}$
\bea
S_{\mathrm{F}}=a^d\sum_{\mathrm{all\ lattice\ points}}\bar{\psi}(D+m_{\mathrm{F}})\psi
\eea
when meeting the conditions:
Dirac matrix $D$ is exponentially local;
loss of massless doublers;
chiral symmetry $\gamma_5D+D\gamma_5=0$.
We could {\it avoid} the no-go by modifying the {\it chiral symmetry} condition (like overlap fermion), but it is necessary to accept the {\it square root} operation with a modified chiral-symmetry.
One way to avoid the ``no-go'' is to introduce an auxiliary term that breaks the chiral symmetry, such as the Wilson-Dirac fermion. 
However, this approach may pose difficulties when studying a light fermion mass. 
The ``no-go'' acts as a strong constraint on the creation of a lattice fermion.
The {\it square-root} operation and {\it losing} chiral symmetry all introduce practical problems about {\it simulation time} or {\it error bar}.
Hence, we consider a naive fermion approach with a one-sided finite-difference. 
This {\it breaks} the hermiticity but does {\it not} violate the no-go \cite{Stamatescu:1993ga, Stamatescu:1994yj}.
Because the one-sided lattice derivative breaks the hypercubic symmetry \cite{Stamatescu:1993ga}, we may not obtain the expected continuum field theory due to the issue of non-renormalizability \cite{Sadooghi:1996ip}.
The imposition of the average over all possible one-sided finite-differences \cite{Stamatescu:1993ga} helps to eliminate the non-renormalizable terms \cite{Sadooghi:1996ip} and allows for the recovery of the continuum field theory. 
However, the non-hermitian theory generates a numerical sign problem, which can cause the MC method to lose the powerful prediction or the importance sampling technique from the sign problem. 
Fortunately, this issue has been resolved by considering the even flavors with the same numbers between the fermion fields with the forward and backward finite-differences \cite{Guo:2021sjp}. 
It is important to note that while the Lagrangian may not be hermitian, the partition function is real \cite{Guo:2021sjp}. 
Therefore, we can avoid the sign problem and implement the Hybrid Monte Carlo (HMC) method in non-hermitian lattice fermion field theory \cite{Guo:2021sjp}.
\\

\noindent
In this paper, we implement HMC on the non-Hermitian lattice formulation for a 2D Gross-Neveu-Yukawa (GNY) model with two flavors. 
The Lagrangian description is 
\bea
S_{\mathrm{GNY}2}\lbrack\bar{\psi}, \psi, \phi\rbrack
=
\int d^2x\ \bigg(\sum_{j=1}^2\big(\bar{\psi}_j(\fsl{\partial}+m_{\mathrm{F}}+\phi)\psi_j\big)-\phi\Box\phi+\frac{1}{2g^2}\phi^2\bigg).
\eea 
This model is interesting because it does not suffer from UV divergence issues, making it suitable for exploring non-Hermitian lattice methods. 
The 2D GNY model with two flavors is utilized due to its ability to handle all eigenmodes of the Dirac matrix, making it a suitable testbed for the non-Hermitian lattice method. 
We validate our numerical solutions by calculating correlators \cite{Schwinger:1951ex} and comparing them with one-loop resummation, particularly in the non-weak coupling regime. 
Because the GNY model has a three-point coupling similar to $\phi^3$ theory, we can use the resummation method of Ref. \cite{Ma:2023uar}. 
The resummation method involves an adaptive parameter determined by minimizing energy expectation values, leading to a modified unperturbed part and a spectrum closer to the full spectrum \cite{Ma:2022atx}. 
Therefore, we call this method Adaptive Perturbation Method \cite{Weinstein:2005kw,Weinstein:2005kx}. 
This method is used for studying the Renormalization Group (RG) flow \cite{Gell-Mann:1954yli}. 
It adjusts parameters based on system behavior, allowing us to explore aspects like condensation, bare masses, and coupling constants. 
To summarize our results:
\begin{itemize}
\item
We compute $\langle\phi\rangle$, $\langle\phi^2\rangle$, $\langle\mathrm{Tr}(\bar{\psi}_1\psi_1+\bar{\psi}_2\psi_2)/2\rangle$, and $\langle\mathrm{Tr}(\bar{\psi}_1\psi_1+\bar{\psi}_2\psi_2)\phi/2\rangle$ in lattice simulations and compare them with the Adaptive Perturbation Method up to the bare Yukawa coupling constant $\lambda=2$. 
This method's leading order matches well with simulation results for the bare fermion mass $m_{\mathrm{F}}=4$, the bare scalar mass $m_{\mathrm{B}}=1$, and the lattice size $(N_t, N_x)=(16, 32)$, where $N_t$ ($N_x$) is the temporal (spatial) lattice size with the unit lattice spacing ($a=1$). 
\item 
By studying the RG flow. we show that the bare Yukawa coupling constant converges to a finite value as the momentum cutoff $\Lambda$ increases, indicating asymptotic safety in the 2D GNY model. 
This insight can be valuable for understanding interacting theories in the continuum limit. 
\end{itemize}

\noindent
The organization of this paper is as follows: 
We start with the discussion of the GNY model on a lattice, moving on to the implementation of HMC, and then comparing simulation with the adaptive perturbation method for two flavors in Sec.~\ref{sec:2}. 
We then use the adaptive perturbation method to give the RG flow and show the asymptotic safe for two flavors in Sec.~\ref{sec:3}. 
Finally, we discuss and conclude in Sec.~\ref{sec:4}. 
In Appendix~\ref{app:a}, we have included the analysis of thermalization and auto-correlation for our lattice data. 

\section{GNY Model on Lattice}
\label{sec:2}
\noindent 
We first review the simulation method in non-Hermitian lattice fermion field theory for even flavors \cite{Guo:2021sjp}. 
We then introduce the 2D GNY model on the lattice with the one-sided finite-difference and compare lattice simulation with the Adaptive Perturbation Method to validate and refine computational techniques. 
Analyzing thermalization and auto-correlation time in lattice data, as provided in Appendix~\ref{app:a}, is crucial evidence for understanding how well the simulation results match theoretical expectations and how efficiently the simulations are running.  

\subsection{Simulation Method}
\noindent
We demonstrate the simulation method in a 1D Dirac fermion field with a degenerate mass for two flavors for simplicity \cite{Guo:2021sjp}
\bea
S_{\mathrm{FD}}=a\sum_{n=0}^{N-1}\bigg(\bar{\psi}_1(n)\big(D(n)+m_{\mathrm{F}}\big)\psi_1(n)+
\bar{\psi}_2(n)\big(-D^{\dagger}(n)+m_{\mathrm{F}}\big)\psi_2(n)\bigg),
\eea
where $N$ is the number of Dirac fermion fields.
The derivative operator of $\psi_1$ is defined by the forward finite-difference
\bea
\frac{\psi_1(n+1)-\psi_1(n)}{a}.
\eea
The derivative operator of another field $\psi_2$ is defined using the backward finite-difference.
After integrating out the fermion fields, we obtain a non-negative determinant \cite{Guo:2021sjp}:
\bea
\det(D+m)\det(-D^{\dagger}+m_{\mathrm{F}})
&=&\det(D+m_{\mathrm{F}})\det\big(\gamma_5(-D^{\dagger}+m_{\mathrm{F}})\gamma_5\big)
\nn\\
&=&|\det(D+m_{\mathrm{F}})|^2.
\eea
We then introduce the pseudo-fermion field (bosonic field $\phi_f$) to rewrite the partition function as in the following
\bea
&&
\int {\cal D}\bar{\psi}{\cal D}\psi\ \exp(-S_{\mathrm{FD}})
\nn\\
&\sim&
\int {\cal D}\phi_{f, {\mathrm{R}}}{\cal D}\phi_{f, \mathrm{I}}\ \exp\big(-\phi_f^{\dagger}\big((D+m_{\mathrm{F}})(D^{\dagger}+m_{\mathrm{F}})\big)^{-1}\phi_f\big),
\eea
where
\bea
\phi_f\equiv\phi_{f, \mathrm{R}}+i\phi_{f, \mathrm{I}}.
\eea
Although it is a non-Hermitian fermion field theory, the partition function is real. 
Hence we can implement the HMC algorithm to compute the observables avoiding the sign problem \cite{Guo:2021sjp}.
\\

\noindent
When considering the GNY model, it is necessary to restore the hypercubic symmetry by averaging over 4 possible orientations \cite{Stamatescu:1993ga, Sadooghi:1996ip}.
We consider forward-forward ($++$) and forward-backward ($+-$) 2D finite-difference for $\psi_1$ field and simultaneously consider backward-backward ($--$) and backward-forward ($-+$) finite-difference in $\psi_2$ field.
When using forward-forward (or forward-backward) finite-difference in $\psi_1$, it is the same as using backward-backward (or backward-forward) finite-difference. 
Therefore, to restore hypercubic symmetry, we can only average two orientations. 
As for the scalar field, we always use forward-forward finite-difference on the lattice. 
The fermion and scalar fields satisfy the boundary conditions: 
\bea
&&
\psi_j(t+N_t, x)=-\psi_j(t, x); \ \psi_j(t, x+N_x)=\psi_j(t, x); 
\nn\\
&&
\phi(t+N_t, x)=\phi(t, x); \ \phi(t, x+N_x)=\phi(t, x). 
\eea 

\subsection{Adaptive Perturbation Method}
\noindent 
We do the one-loop resummation as in $\phi^3$ theory \cite{Ma:2023uar}. 
For one flavor case, we obtain the constant condensation
\bea
\phi_0=\frac{1}{\gamma_{\mathrm{B}}^2(0)}\mathrm{Tr}\bigg(\int\frac{d^2p}{(2\pi)^2}\ \lambda_{\mathrm{R}}(p, p)\frac{1}{i\fsl{p}+\gamma_{\mathrm{F}}(p)}\bigg);  
\eea
the renormalized mass parameters:
\bea
\gamma_{\mathrm{F}}(p)&=&m_{\mathrm{F}}+\lambda_{\mathrm{R}}(p, p)\phi_0
-\int\frac{d^2q_1}{(2\pi)^2}\ \lambda_{\mathrm{R}}(p, q_1)\frac{1}{i\fsl{q}_1+\gamma_{\mathrm{F}}(q_1)}\frac{1}{(p-q_1)^2+\gamma_{\mathrm{B}}^2(p-q_1)}\lambda_{\mathrm{R}}(p, q_1),   
\nn\\
\gamma_{\mathrm{B}}^2(p)&=&m_B^2
+\mathrm{Tr}\bigg(\int\frac{d^2q_1}{(2\pi)^2}\ \lambda_{\mathrm{R}}(q_1, q_1-p)
\frac{1}{i(\fsl{p}-\fsl{q}_1)+\gamma_{\mathrm{F}}(p-q_1)}\lambda_{\mathrm{R}}(q_1, q_1-p)
\frac{1}{i\fsl{q}_1+\gamma_{\mathrm{F}}(q_1)}\bigg); 
\nn\\
\eea  
the renormalized coupling constant
\bea
\lambda_{\mathrm{R}}(p_1, p_2)=\lambda+F(p_1, p_2),   
\eea    
where 
\bea
&&
F(p_1, p_2)
\nn\\
&=&\int \frac{d^2q_1}{(2\pi)^2}\ \lambda_{\mathrm{R}}(p_1, p_1-q_1)
\frac{1}{q_1^2+\gamma_{\mathrm{B}}^2(q_1)}
\frac{1}{i(\fsl{p_1}-\fsl{q_1})+\gamma_{\mathrm{F}}(p_1-q_1)}
\nn\\
&&\times
\lambda_{\mathrm{R}}(p_1-q_1, p_2-q_1)
\frac{1}{i(\fsl{p_2}-\fsl{q_1})+\gamma_{\mathrm{F}}(p_2-q_1)}
\lambda_{\mathrm{R}}(p_2-q_1, p_2)
.
\eea 
We note that $\gamma_{\mathrm{F}}(p)$ and $\lambda_{\mathrm{R}}$ are matrices, not scalars. 
Our notation for the gamma matrices is: 
 \bea
 \gamma_1\equiv\sigma_x=\begin{pmatrix}
 0&1
 \\
 1&0
 \end{pmatrix}, \ 
 \gamma_2\equiv-\sigma_y=\begin{pmatrix}
 0&i
 \\
 -i&0 
 \end{pmatrix}. 
 \eea

\subsection{Comparison} 
\noindent
To compare the lattice simulation with the adaptive perturbation method, we consider two flavors with degenerate fermion masses.
After adopting the discretization with the infinite lattice size limit and the unit lattice spacing, the formula of propagators becomes:
\bea
\int\frac{d^2p}{(2\pi)^2}\ \frac{1}{i\fsl{p}+\gamma_{\mathrm{F}}(p)}\rightarrow
\int\frac{d^2p}{(2\pi)^2}\ \frac{1}{\sum_{\mu=1}^2\epsilon_{\mu}\gamma_{\mu}(\exp(i\epsilon_{\mu}p_{\mu})-1)
+\gamma_{\mathrm{F}}(p)};
\eea
\bea
\int\frac{d^2p}{(2\pi)^2}\ \frac{1}{p^2+\gamma^2_{\mathrm{B}}(p)}\rightarrow
\int\frac{d^2p}{(2\pi)^2}\ \frac{1}{\gamma^2_{\mathrm{B}}(p)+\sum_{\mu=1}^2\big(2-2\cos(p_{\mu})\big)}.
\eea
We can use a similar way to obtain the correlation function on the lattice and use the result to compare it with the lattice simulation.
$\epsilon_{\mu}$ can be 1 or -1 for each $\mu$.
We only need to do the quenched averaging for two choices, $(\epsilon_1, \epsilon_2)=(1, 1)$ and $(\epsilon_1, \epsilon_2)=(1, -1)$, because another two choices give the same result.
\\

\noindent
To simulate the one-sided lattice fermion theory without the sign problem, we use distinct finite-difference operators for $\psi_1$ and $\psi_2$.
It is possible for the renormalized fermion masses, $\gamma_{\mathrm{F}, 1}$ and $\gamma_{\mathrm{F}, 2}$, as well as coupling constants, $\lambda_{\mathrm{R}, 1}$ and $\lambda_{\mathrm{R}, 2}$, to differ. However, our numerical solutions have shown that the matrices have the same real eigenvalues across different fermion fields, which is an interesting observation.
The constant condensation is given by
\bea
\phi_0&=&\frac{1}{\gamma_{\mathrm{B}}^2(0)}\mathrm{Tr}\bigg(\int\frac{d^2p}{(2\pi)^2}\ \lambda_{\mathrm{R}, 1}(p, p) \frac{1}{\sum_{\mu=1}^2\epsilon_{1,\mu}\gamma_{\mu}(\exp(i\epsilon_{1,\mu}p_{\mu})-1) +\gamma_{\mathrm{F}, 1}(p)} \bigg)
\nn\\
&&+ \frac{1}{\gamma_\mathrm{B}^2(0)}\mathrm{Tr}\bigg(\int\frac{d^2p}{(2\pi)^2}\ \lambda_{\mathrm{R}, 2}(p, p) \frac{1}{\sum_{\mu=1}^2\epsilon_{2,\mu}\gamma_{\mu}(\exp(i\epsilon_{2,\mu}p_{\mu})-1) +\gamma_{\mathrm{F}, 2}(p)} \bigg);
\nn\\
\eea
the renormalized fermion masses are:
\bea
\gamma_{\mathrm{F},1}(p)&=&m_{\mathrm{F}}+\lambda_{\mathrm{R},1}(p, p)\phi_0
\nn\\
&&- \int\frac{d^2q_1}{(2\pi)^2}\ \lambda_{\mathrm{R}, 1}(p, q_1) \frac{1}{\sum_{\mu=1}^2\epsilon_{1,\mu}\gamma_{\mu}(\exp(i\epsilon_{1,\mu}q_{1, \mu})-1) +\gamma_{\mathrm{F}, 1}(p)}
\nn\\
&&\times
\frac{1}{\gamma^2_{\mathrm{B}}(p-q_1)+\sum_{\mu=1}^2\big(2-2\cos(p_{\mu}-q_{1, \mu})\big)}
\lambda_{\mathrm{R}, 1}(p, q_1),
\nn\\
\gamma_{\mathrm{F}, 2}(p)&=&m_{\mathrm{F}}+\lambda_{\mathrm{R},2}(p, p)\phi_0
\nn\\
&&- \int\frac{d^2q_1}{(2\pi)^2}\ \lambda_{\mathrm{R}, 2}(p, q_1) \frac{1}{\sum_{\mu=1}^2\epsilon_{1,\mu}\gamma_{\mu}(\exp(i\epsilon_{1,\mu}q_{1, \mu})-1) +\gamma_{\mathrm{F},  2}(p)}
\nn\\
&&\times
\frac{1}{\gamma^2_{\mathrm{B}}(p-q_1)+\sum_{\mu=1}^2\big(2-2\cos(p_{\mu}-q_{1, \mu})\big)}
\lambda_{\mathrm{R}, 2}(p, q_1);
\eea
the square of renormalized scalar mass is
\bea
\gamma_\mathrm{B}^2&=&m_\mathrm{B}^2
\nn\\
&&+\mathrm{Tr}\bigg(\int\frac{d^2q_1}{(2\pi)^2}\ \lambda_{\mathrm{R}, 1}(p, q_1) \frac{1}{\sum_{\mu=1}^2\epsilon_{1,\mu}\gamma_{\mu}(\exp(i\epsilon_{1,\mu}(p-q_{1, \mu}))-1) +\gamma_{F,1}(p-q_1)}
\nn\\
&&\times
\lambda_{\mathrm{R}, 1}(q_1, q_1-p)
\frac{1}{\sum_{\mu=1}^2\epsilon_{1,\mu}\gamma_{\mu}(\exp(i\epsilon_{1,\mu}q_{1, \mu})-1) +\gamma_{\mathrm{F}, 1}(q_1)} \bigg)
\nn\\
&&+\mathrm{Tr}\bigg(\int\frac{d^2q_1}{(2\pi)^2}\ \lambda_{\mathrm{R}, 2}(p, q_1) \frac{1}{\sum_{\mu=1}^2\epsilon_{1,\mu}\gamma_{\mu}(\exp(i\epsilon_{1,\mu}(p-q_{1, \mu}))-1) +\gamma_{F,1}(p-q_1)}
\nn\\
&&\times
\lambda_{\mathrm{R}, 2}(q_1, q_1-p)
\frac{1}{\sum_{\mu=1}^2\epsilon_{1,\mu}\gamma_{\mu}(\exp(i\epsilon_{1,\mu}q_{1, \mu})-1) +\gamma_{\mathrm{F}, 1}(q_1)} \bigg);
\eea
the renormalized coupling constants are:
\bea
\lambda_{\mathrm{R}, 1}(p_1, p_2)=\lambda+F_1(p_1, p_2), \ \lambda_{\mathrm{R}, 2}(p_1, p_2)=\lambda+F_2(p_1, p_2),
\eea
where
\bea
F_1(p_1, p_2) &=&
\int \frac{d^2q_1}{(2\pi)^2}\ \lambda_{\mathrm{R}, 1}(p_1, p_1-q_1)
\frac{1}{\gamma^2_{\mathrm{B}}(q_1)+\sum_{\mu=1}^2\big(2-2\cos(q_{1, \mu})\big)}
\nn\\
&&\times
\frac{1}{\sum_{\mu=1}^2\epsilon_{1,\mu}\gamma_{\mu}\bigg(\exp\big(i\epsilon_{1,\mu}(p_{\mu, 1}-q_{\mu, 1})\big)-1\bigg) +\gamma_{\mathrm{F}, 1}(p_1-q_1)}
\nn\\
&&\times
\lambda_{\mathrm{R}, 1}(p_1-q_1, p_2-q_1)
\nn\\
&&\times
\frac{1}{\sum_{\mu}\epsilon_{1,\mu}\gamma_{\mu}\bigg(\exp\big(i\epsilon_{1,\mu}(p_{\mu, 2}-q_{\mu, 1})\big)-1\bigg) +\gamma_{\mathrm{F}, 1}(p_2-q_1)}
\nn\\
&&\times
\lambda_{\mathrm{R}, 1}(p_2-q_1, p_2),
\nn\\
F_2(p_1, p_2) &=&
\int \frac{d^2q_1}{(2\pi)^2}\ \lambda_{\mathrm{R}, 2}(p_1, p_1-q_1)
\frac{1}{\gamma^2_{\mathrm{B}}(q_1)+\sum_{\mu=1}^2\big(2-2\cos(q_{1, \mu})\big)}
\nn\\
&&\times
\frac{1}{\sum_{\mu=1}^2\epsilon_{1,\mu}\gamma_{\mu}\bigg(\exp\big(i\epsilon_{1,\mu}(p_{\mu, 1}-q_{\mu, 1})\big)-1\bigg) +\gamma_{\mathrm{F}, 2}(p_1-q_1)}
\nn\\
&&\times
\lambda_{\mathrm{R}, 2}(p_1-q_1, p_2-q_1)
\nn\\
&&\times
\frac{1}{\sum_{\mu}\epsilon_{1,\mu}\gamma_{\mu}\bigg(\exp\big(i\epsilon_{1,\mu}(p_{\mu, 2}-q_{\mu, 1})\big)-1\bigg) +\gamma_{\mathrm{F}, 2}(p_2-q_1)}
\nn\\
&&\times
\lambda_{\mathrm{R}, 2}(p_2-q_1, p_2).
\eea
\\

\begin{figure}
\includegraphics[width=0.49\textwidth]{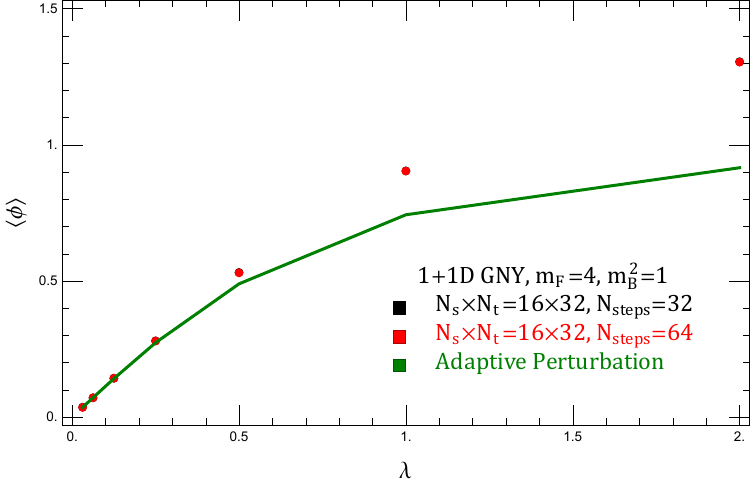}
\includegraphics[width=0.49\textwidth]{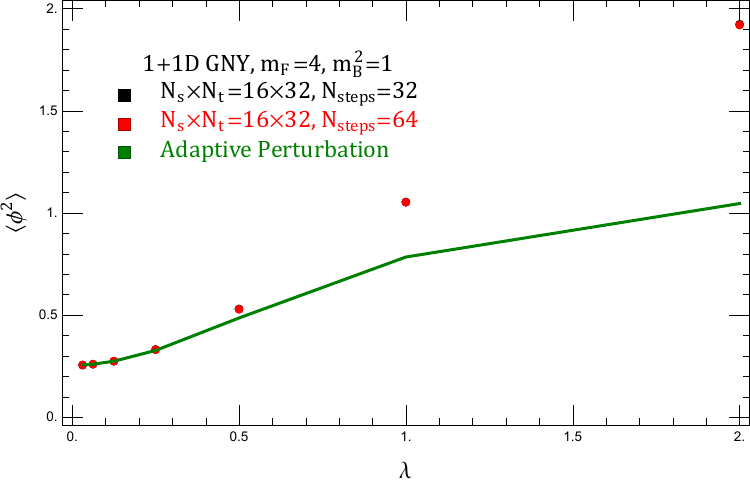}
\includegraphics[width=0.49\textwidth]{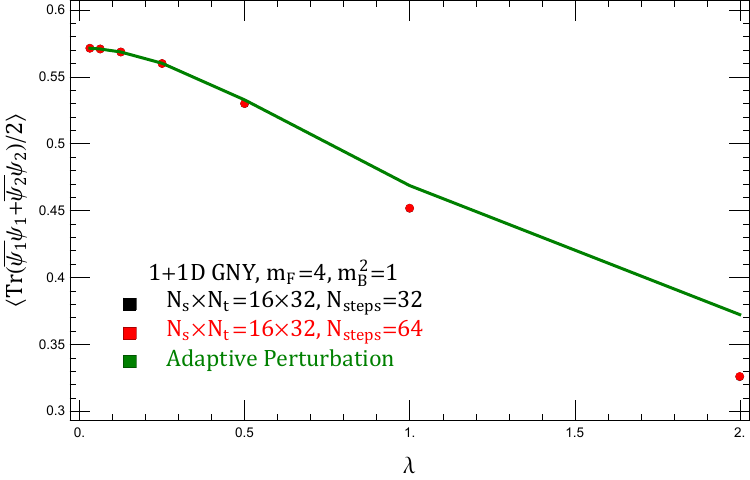}
\includegraphics[width=0.49\textwidth]{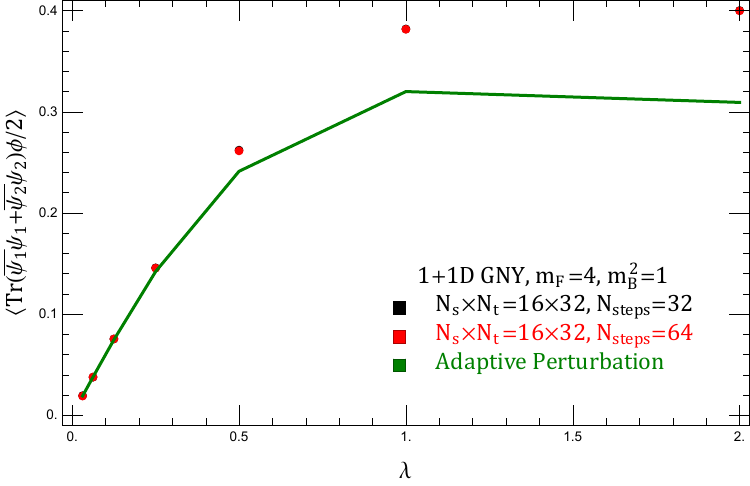}
\caption{We compare the perturbation result to the HMC simulation in $\langle\phi\rangle$, $\langle\phi^2\rangle$, $\langle\mathrm{Tr}(\bar{\psi}_1\psi_1+\bar{\psi}_2\psi_2)/2\rangle$, and $\langle\mathrm{Tr}(\bar{\psi}_1\psi_1+\bar{\psi}_2\psi_2)\phi/2\rangle$ for $m_{\mathrm{F}}=4$ and $m^2_{\mathrm{B}}=1$ with the lattice size $(N_s, N_t)=(16, 32)$ and the number of molecular dynamics steps $N_{\mathrm{steps}}=32, 64$.
The number of measurements is $2^{11}$ sweeps with thermalization $2^6$ sweeps and measure intervals $2^5$ sweeps.
}
\label{comparison}
\end{figure}

\begin{figure}
\includegraphics[width=0.49\textwidth]{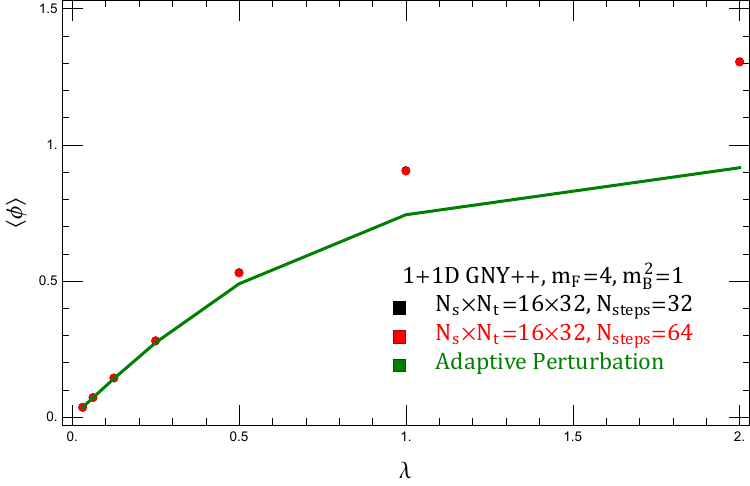}
\includegraphics[width=0.49\textwidth]{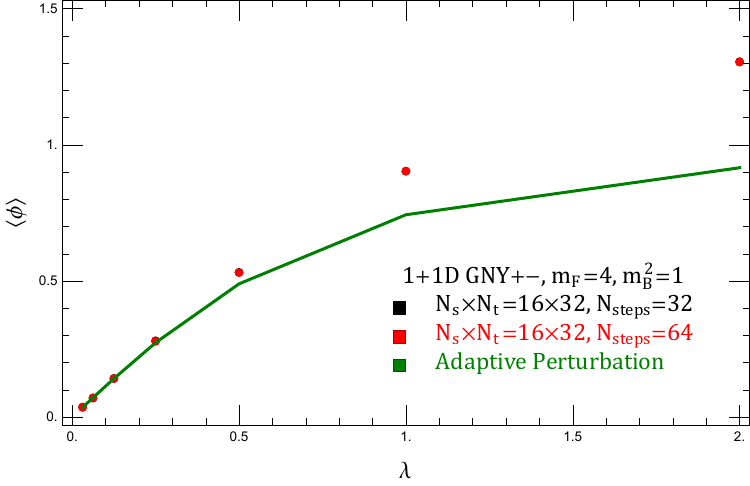}
\includegraphics[width=0.49\textwidth]{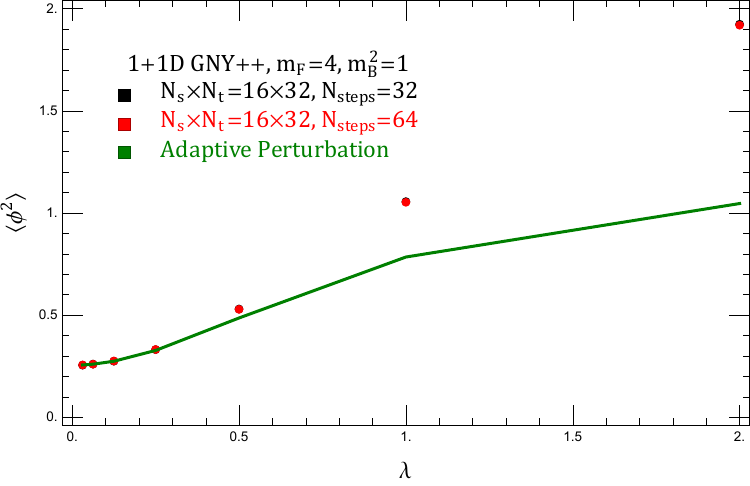}
\includegraphics[width=0.49\textwidth]{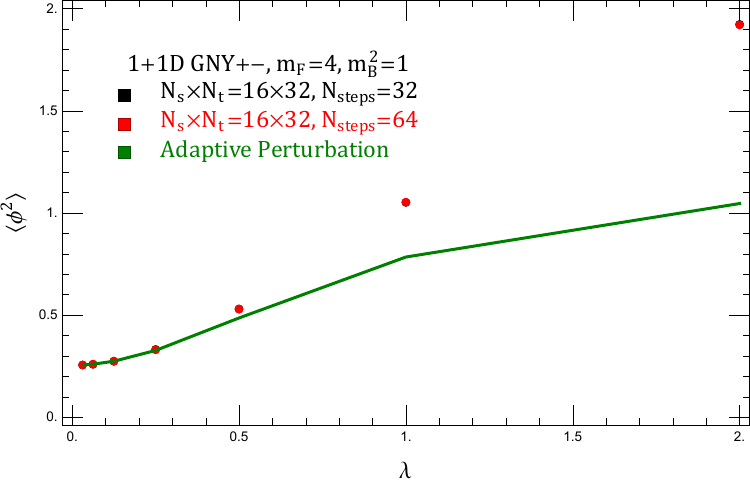}
\includegraphics[width=0.49\textwidth]{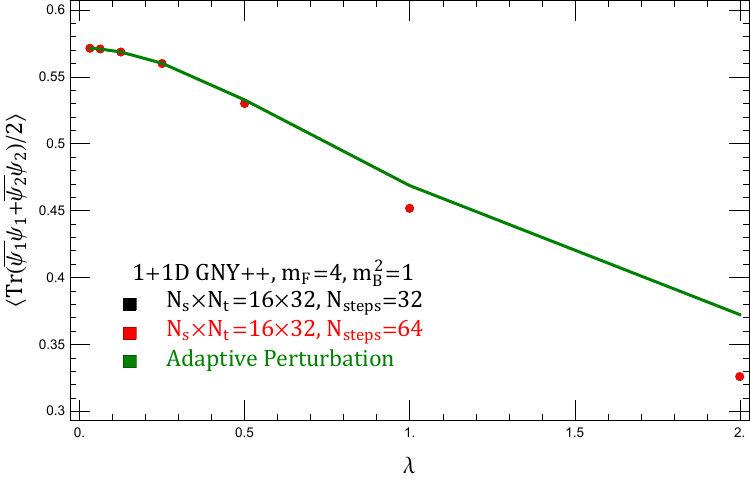}
\includegraphics[width=0.49\textwidth]{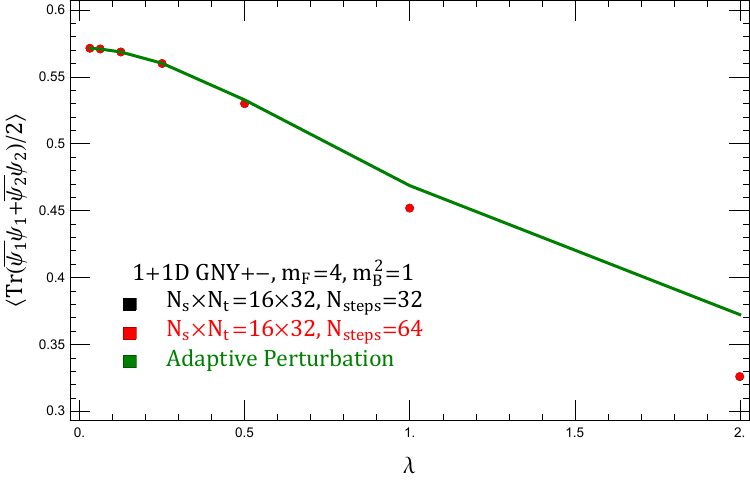}
\includegraphics[width=0.49\textwidth]{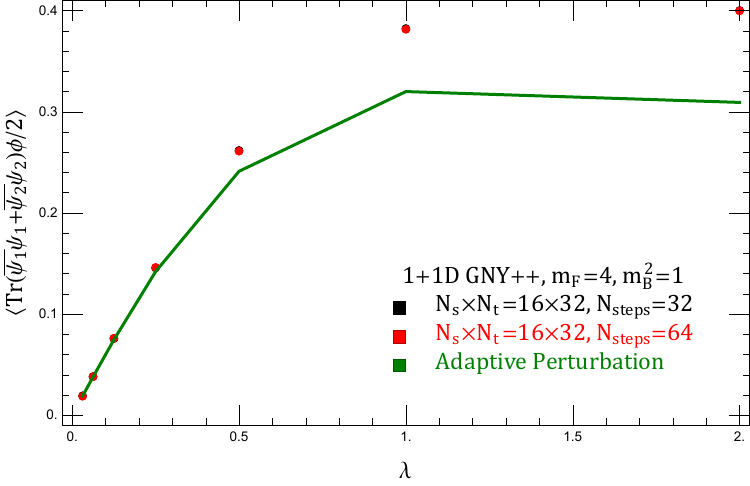}
\includegraphics[width=0.49\textwidth]{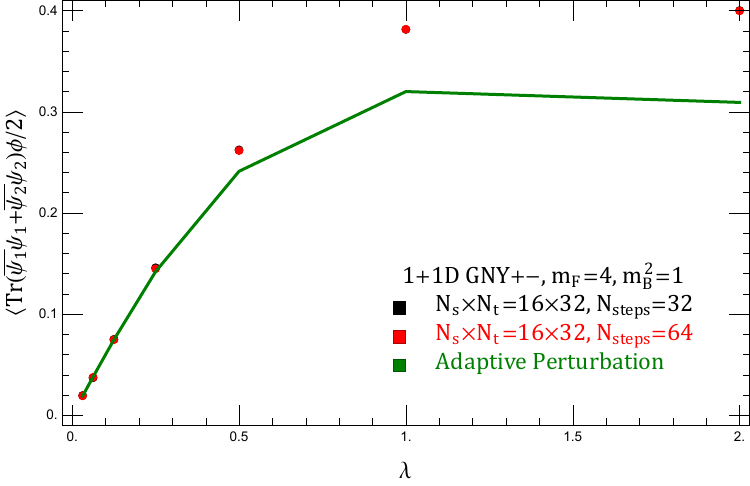}
\caption{We compare the perturbation result to the HMC simulation in $\langle\phi\rangle$, $\langle\phi^2\rangle$, $\langle\mathrm{Tr}(\bar{\psi}_1\psi_1+\bar{\psi}_2\psi_2)/2\rangle$, and $\langle\mathrm{Tr}(\bar{\psi}_1\psi_1+\bar{\psi}_2\psi_2)\phi/2\rangle$ without the quenched averaging for $m_{\mathrm{F}}=4$ and $m^2_{\mathrm{B}}=1$ with the lattice size $(N_s, N_t)=(16, 32)$ and the number of molecular dynamics steps $N_{\mathrm{steps}}=32, 64$.
The number of measurements is $2^{11}$ sweeps with thermalization $2^6$ sweeps and measure intervals $2^5$ sweeps.
}
\label{comparison+++-}
\end{figure}

\noindent
We can use the condensation and the renormalized parameters to write the leading-order result of the adaptive perturbation method and compare the result to the lattice simulation in Fig. \ref{comparison}.
The difference between the numerical solution and the perturbation solution becomes clear at $\lambda=0.5$.
However, the difference is still within 10\%, and the next-order perturbation correction of $\langle\phi\rangle$ is positive.
Therefore, we expect that the difference is due to the loss of the higher-order correction of the perturbation solution.
We also show the comparison before averaging the orientations in Fig. \ref{comparison+++-}.
The result is similar to the quenched averaging, Fig. \ref{comparison}.

\section{RG Flow}
\label{sec:3}
\begin{figure}
\includegraphics[width=0.32\textwidth]{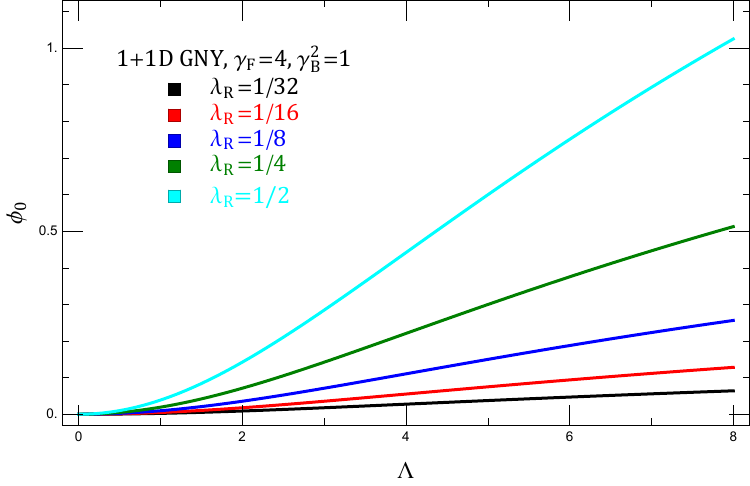}
\includegraphics[width=0.32\textwidth]{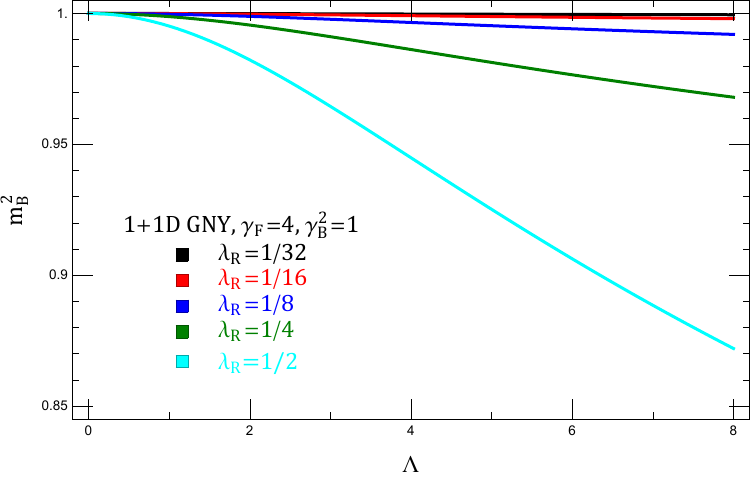}
\includegraphics[width=0.32\textwidth]{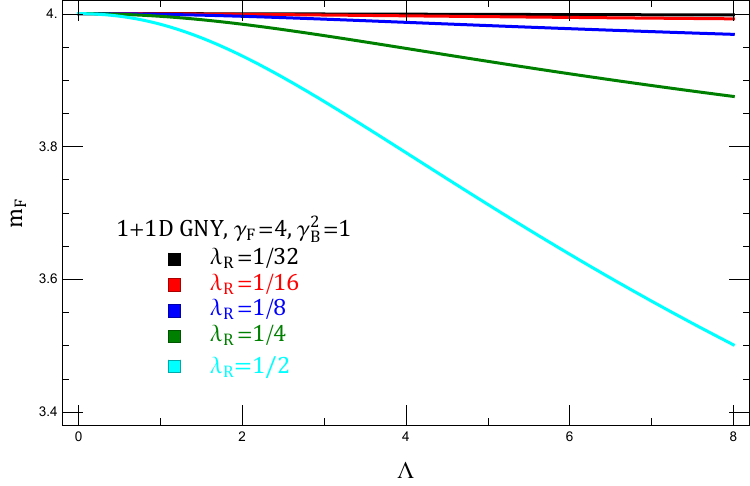}
\caption{We show the RG flow of the condensation, bare scalar mass, and bare fermion mass for the $\gamma_{\mathrm{F}}=4$, $\gamma^2_{\mathrm{B}}=1$, and $\lambda_{\mathrm{R}}=$1/32, 1/16, 1/8, 1/4, 1/2 from the Adaptive Perturbation Method.
}
\label{RGflow1}
\end{figure}

\begin{figure}
\includegraphics[width=0.49\textwidth]{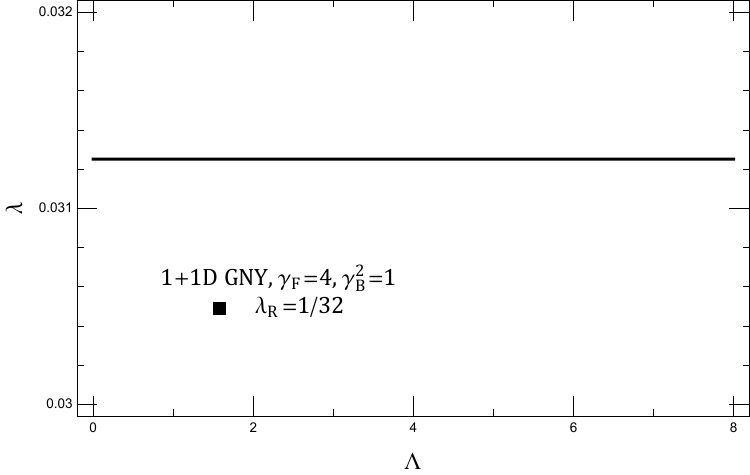}
\includegraphics[width=0.49\textwidth]{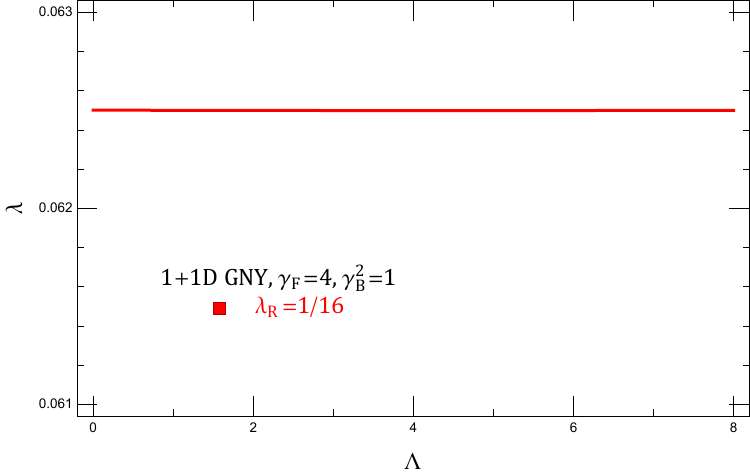}
\includegraphics[width=0.49\textwidth]{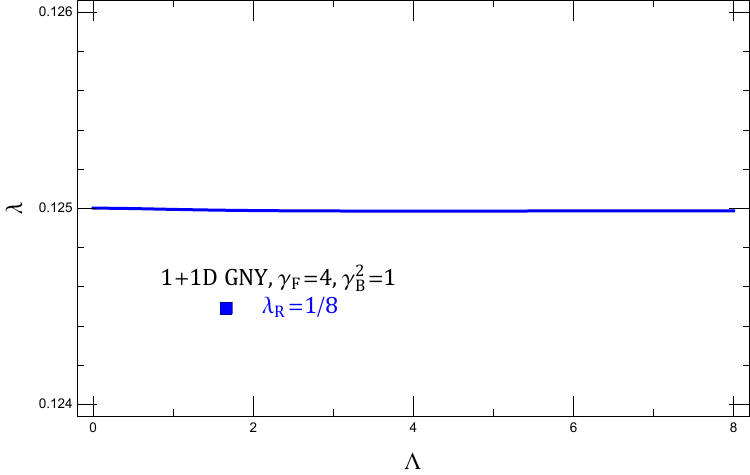}
\includegraphics[width=0.49\textwidth]{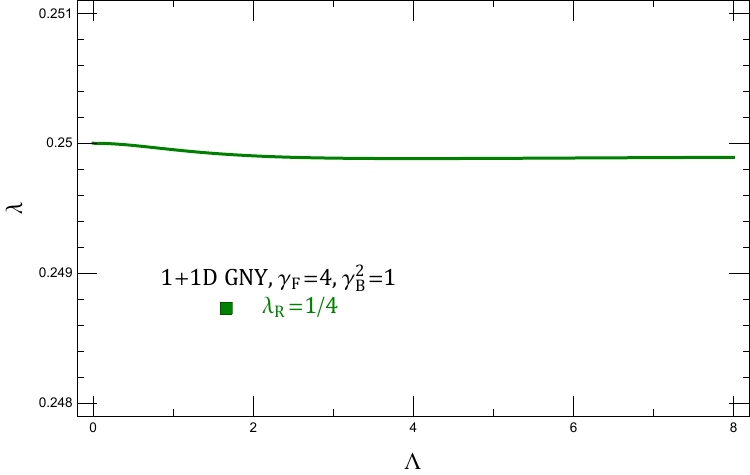}
\includegraphics[width=0.49\textwidth]{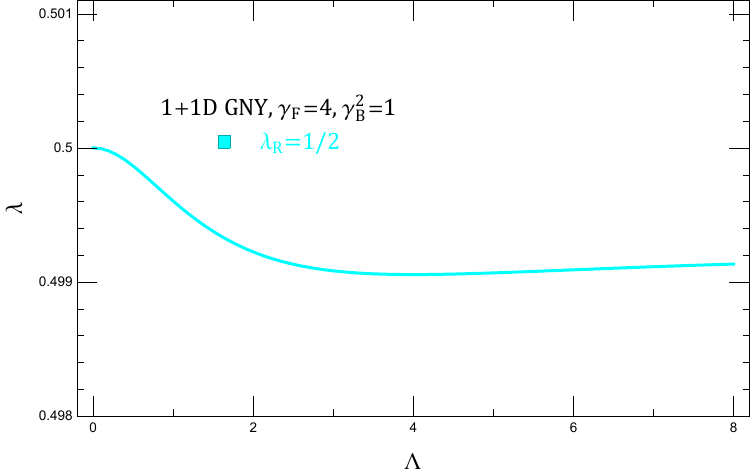}
\caption{We show the RG flow of the bare coupling constant for the $\gamma_{\mathrm{F}}=4$, $\gamma^2_{\mathrm{B}}=1$, and $\lambda_{\mathrm{R}}=$1/32, 1/16, 1/8, 1/4, 1/2 from the Adaptive Perturbation Method.
The bare coupling constant approaches a finite constant as $\Lambda$ increases.
Therefore, the 2D GNY model is asymptotic safety.
}
\label{RGflow2}
\end{figure}

\noindent 
We use the leading-order result of the Adaptive Perturbation Method to obtain the RG flow of the constant condensation, bare scalar mass, bare fermion mass, and bare coupling constant.
At the leading-order result, we can identify the renormalized parameters as physical parameters when the square of momentum equals the negative square of physical mass. 
This means that these parameters can now be used to make predictions about observable phenomena. 
The bare parameters at $\Lambda=0$ are identified as the renormalized parameters. 
For simplicity, we denote $\gamma_{\mathrm{B}}$, $\gamma_{\mathrm{F}}$, and $\lambda_{\mathrm{R}}$ as the physical parameters without writing the momentum dependence in this section.  
We assume that $\gamma_{\mathrm{F}}$ and $\lambda_{\mathrm{R}}$ are proportional to an identity matrix. 
The flow of the condensation, bare scalar mass, and fermion mass is illustrated in Fig. \ref{RGflow1}. 
We show that the 2D GNY model is asymptotic safety from the RG flow of the bare coupling constant (Fig. \ref{RGflow2}).

\newpage
\section{Discussion and Conclusion}
\label{sec:4}
\noindent
We implemented HMC on the non-Hermitian lattice interacting fermion fields within a 2D GNY model with two flavors.
It is the first lattice simulation for the interacting field theory with a large lattice size $(N_t, N_x)=(16, 32)$.
Comparing our numerical results with the Adaptive Perturbation Method up to a coupling constant of $\lambda=2$ is a thorough approach to validation.
Furthermore, utilizing the adaptive perturbation method to derive the RG flow and demonstrate asymptotic safety.
This suggests that the 2D GNY model with two flavors avoids UV divergence.
The interaction that persists in the lattice formulation even at the continuum limit is intriguing.
Our research could have significant implications for understanding the behavior of fermion fields in non-Hermirian lattice formulation, particularly within the context of lattice simulations and QFT.
\\ 

\noindent 
The study of lattice fermion field theory, particularly in the context of the GNY model, offers fascinating insights into QFT and its applications in various dimensions.
Our approach to not truncating the eigenmodes of the Dirac matrix in the 2D lattice fermion field theory is intriguing.
By retaining all eigenmodes, we are likely aiming for a more accurate representation of the system's behavior, albeit at the cost of increased computational complexity.
The GNY model's property of being renormalizable in both 2D and 4D makes it particularly valuable for theoretical investigations.
It provides a platform for studying fundamental aspects of QFT, such as renormalization and phase transitions, in different dimensionalities.
We point out that simulating 4D lattice fermion fields within a reasonable time frame by considering only low-lying eigenmodes is crucial for practical computational purposes.
This approach allows researchers to focus computational resources on the most relevant aspects of the system, potentially paving the way for more efficient simulations and deeper insights into the behavior of higher-dimensional systems.
Indeed, utilizing the 2D GNY model as a laboratory for testing truncation errors before extending the analysis to the 4D GNY model is a wise strategy.
It enables researchers to refine their computational techniques and gain confidence in the validity of their results before tackling more challenging and resource-intensive simulations.
\\

\noindent
We discuss theoretical physics concepts related to lattice models, specifically the GNY model and its similarities to QCD in terms of phase diagrams, the sign problem, and analytical continuation techniques. 
The lattice fermion fields with even flavors and imaginary chemical potentials being free from the sign problem is an interesting observation, as it suggests a potential workaround for numerical simulations. 
Analytical continuation is indeed a powerful tool in theoretical physics, allowing researchers to infer real-chemical potential behavior from calculations done in more tractable regimes, such as imaginary chemical potentials. 
\\

\noindent
The comparison between analytical and numerical results, as we suggested, is crucial for validating theoretical approaches and numerical methods.
If the results align, it strengthens confidence in both the theoretical framework and the computational techniques used.
This process can shed light on the phase diagram of the system and potentially uncover features like the tri-critical point.
Developing new tools to probe such intricate phenomena is at the forefront of theoretical physics research, especially in fields like QFT and condensed matter physics.
The interplay between theoretical insights, numerical simulations, and experimental data is vital for advancing our understanding of complex systems.

\section*{Acknowledgments} 
\noindent
We would like to express our gratitude to Sinya Aoki for his helpful discussion. 
CTM would like to thank Nan-Peng Ma for his encouragement.
XG acknowledges the Guangdong Major Project of Basic and Applied Basic Research (Grant No. 2020B0301030008) and the National Natural Science Foundation of China (Grant No.11905066).
CTM acknowledges the Nuclear Physics Quantum Horizons program through the Early Career Award (Grant No. DE-SC0021892);
YST Program of the APCTP;
China Postdoctoral Science Foundation, Postdoctoral General Funding: Second Class (Grant No. 2019M652926).
HZ acknowledges the Guangdong Major Project of Basic and Applied Basic Research (Grant No. 2020B0301030008);
the Science and Technology Program of Guangzhou (Grant No. 2019050001);
the National Natural Science Foundation of China (Grant Nos. 12047523 and 12105107).

\appendix
\section{Analysis of Thermalization and Auto-Correlation Time} 
\label{app:a}
\noindent
We discuss a scientific analysis of thermalization from the time-history and auto-correlation time from the error bar. 

\subsection{Thermalization} 
\begin{figure}
\includegraphics[width=0.49\textwidth]{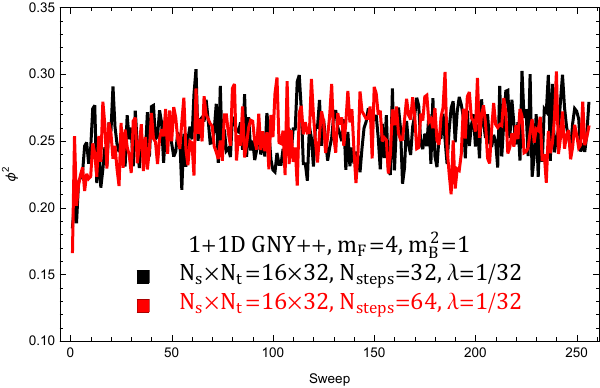}
\includegraphics[width=0.49\textwidth]{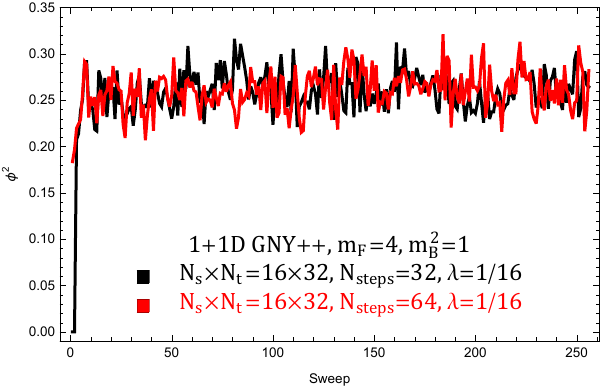}
\includegraphics[width=0.49\textwidth]{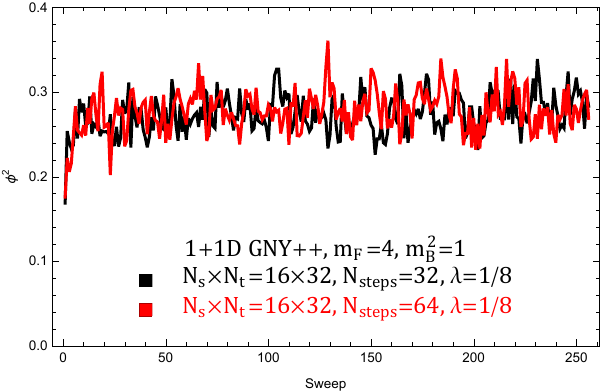}
\includegraphics[width=0.49\textwidth]{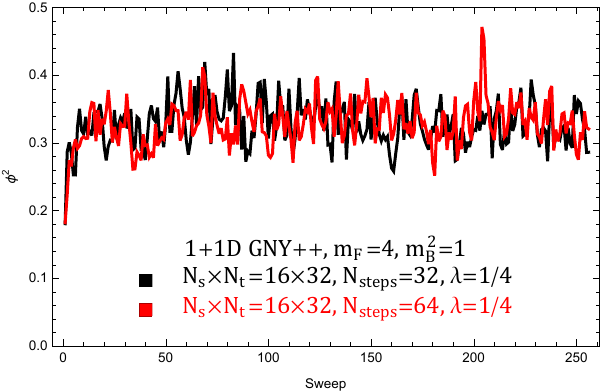}
\includegraphics[width=0.49\textwidth]{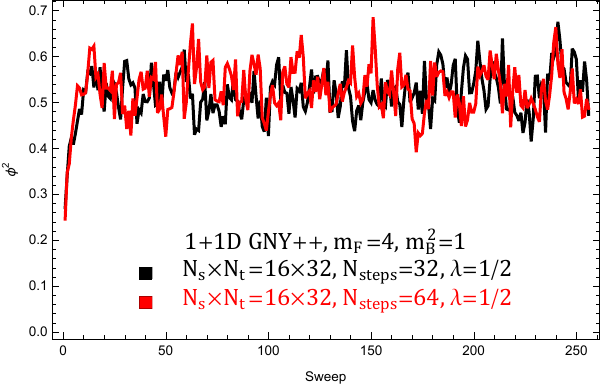}
\includegraphics[width=0.49\textwidth]{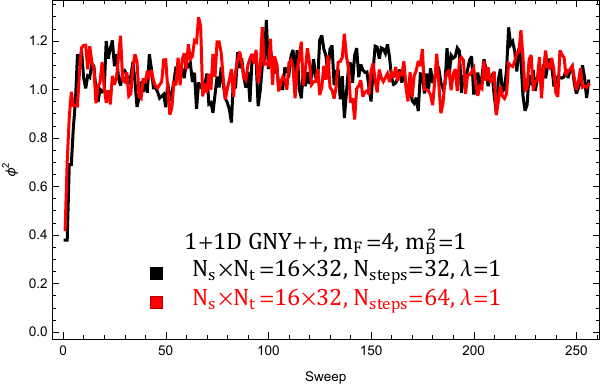}
\includegraphics[width=0.49\textwidth]{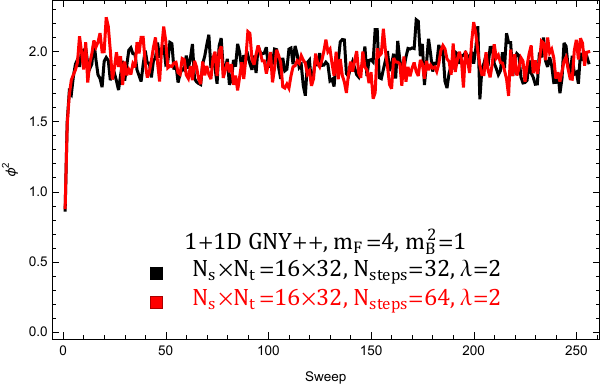}
\caption{The time-history of $\phi^2$ for $m_{\mathrm{F}}=4$, $m^2_{\mathrm{B}}=1$, and $\lambda=1/32, /16, 1/8, 1/4, 1/2, 1, 2$ with the lattice size $(N_s, N_t)=(16, 32)$ and the number of molecular dynamics steps $N_{\mathrm{steps}}=32, 64$ in the 2D GNY model with the forward-forward finite-difference.
}
\label{TH1}
\end{figure}

\begin{figure}
\includegraphics[width=0.49\textwidth]{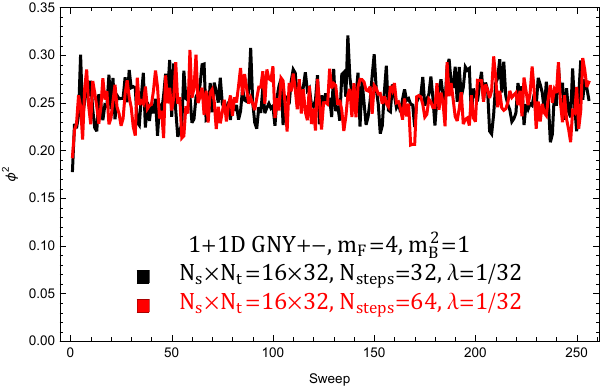}
\includegraphics[width=0.49\textwidth]{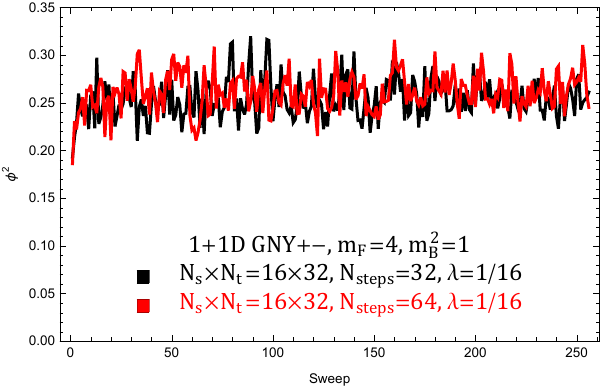}
\includegraphics[width=0.49\textwidth]{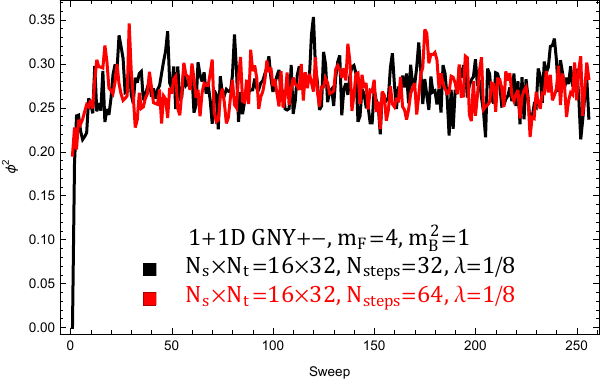}
\includegraphics[width=0.49\textwidth]{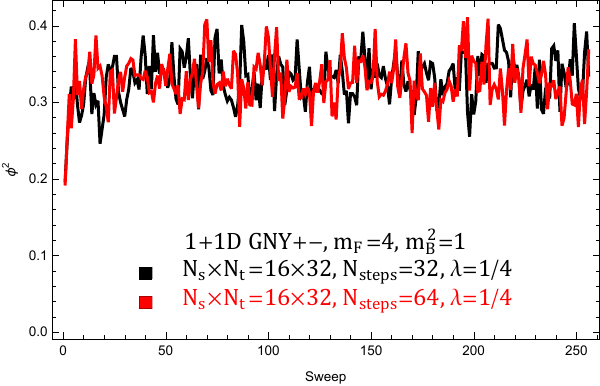}
\includegraphics[width=0.49\textwidth]{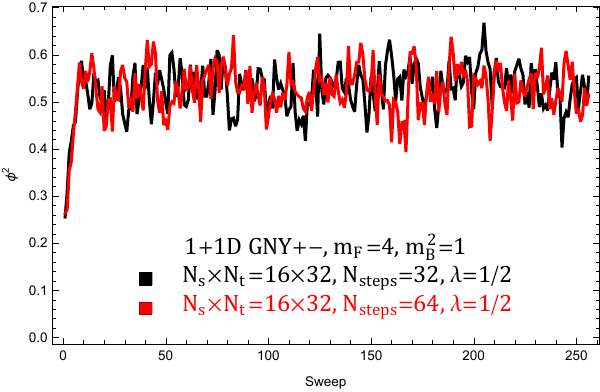}
\includegraphics[width=0.49\textwidth]{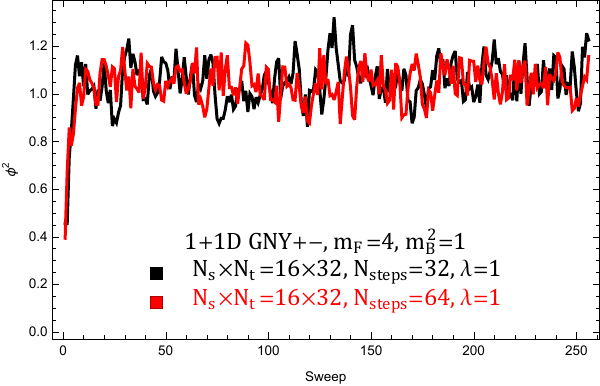}
\includegraphics[width=0.49\textwidth]{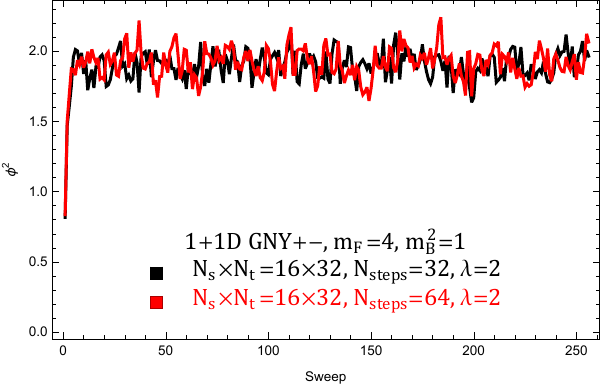}
\caption{The time-history of $\phi^2$ for $m_{\mathrm{F}}=4$, $m^2_{\mathrm{B}}=1$, and $\lambda=1/32, /16, 1/8, 1/4, 1/2, 1, 2$ with the lattice size $(N_s, N_t)=(16, 32)$ and the number of molecular dynamics steps $N_{\mathrm{steps}}=32, 64$ in the 2D GNY model with the forward-backward finite-difference.
}
\label{TH2}
\end{figure}

\noindent 
We show the time history of $\phi^2$ without the quenched averaging in Figs. \ref{TH1} and \ref{TH2}.
The time history justifies the approximate thermalization after 64 weeps.

\subsection{Auto-Correlation Time} 
\begin{figure}
\includegraphics[width=0.49\textwidth]{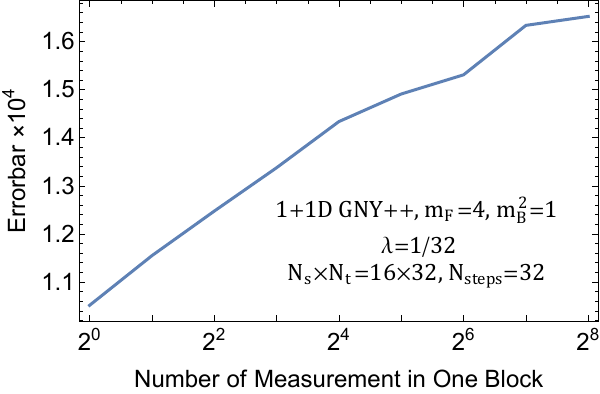}
\includegraphics[width=0.49\textwidth]{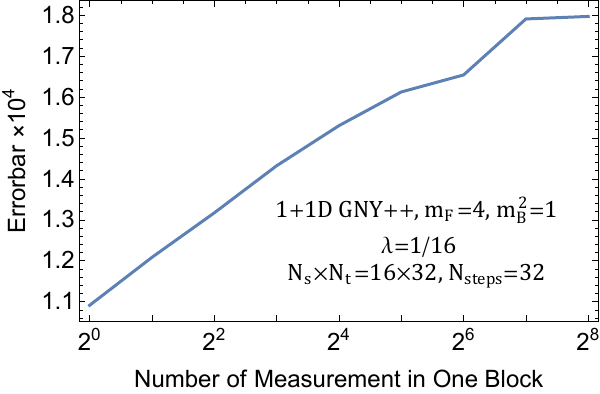}
\includegraphics[width=0.49\textwidth]{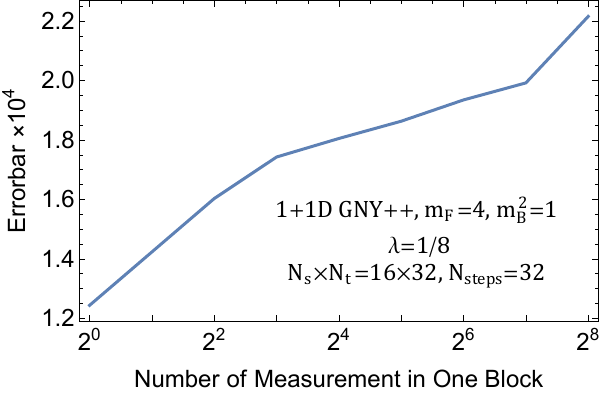}
\includegraphics[width=0.49\textwidth]{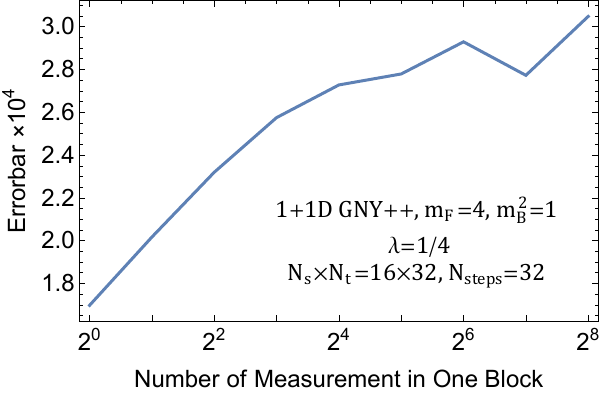}
\includegraphics[width=0.49\textwidth]{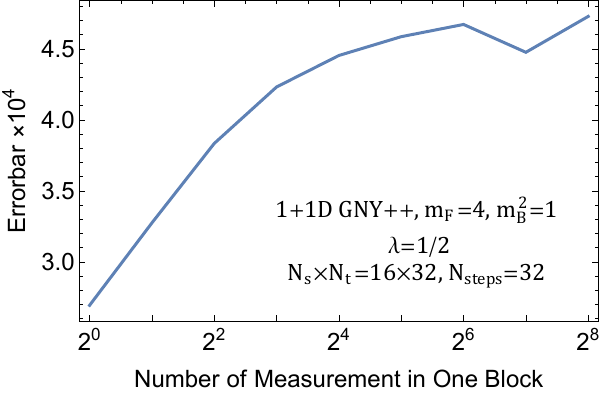}
\includegraphics[width=0.49\textwidth]{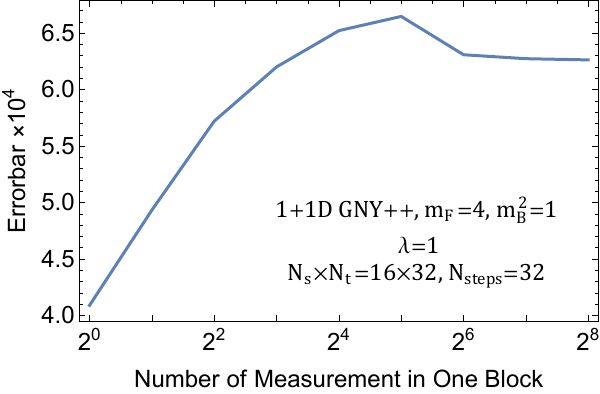}
\includegraphics[width=0.49\textwidth]{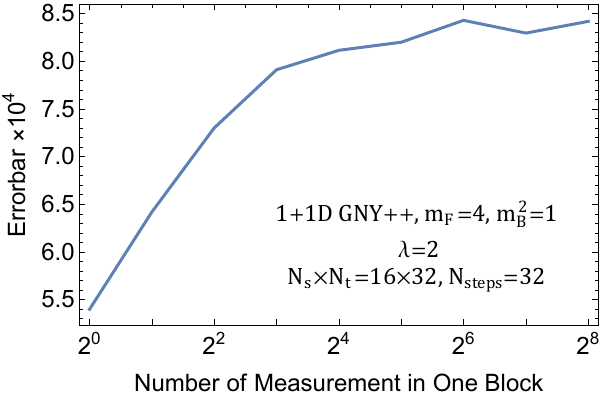}
\caption{The error bar of $\phi^2$ for $m_{\mathrm{F}}=4$, $m^2_{\mathrm{B}}=1$, and $\lambda=1/32, /16, 1/8, 1/4, 1/2, 1, 2$ with the lattice size $(N_s, N_t)=(16, 32)$ and the number of molecular dynamics steps $N_{\mathrm{steps}}=32$ in the 2D GNY model with the forward-forward finite-difference.
}
\label{EB1}
\end{figure}

\begin{figure}
\includegraphics[width=0.49\textwidth]{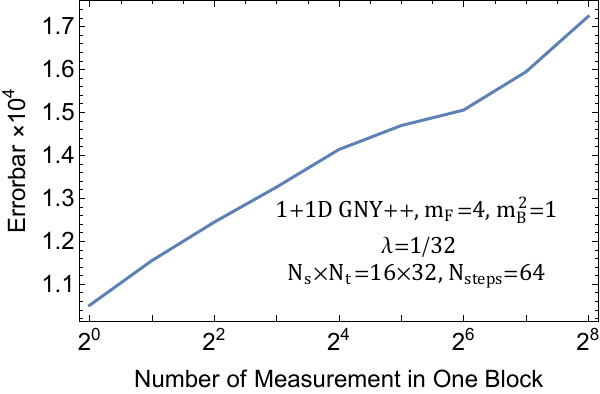}
\includegraphics[width=0.49\textwidth]{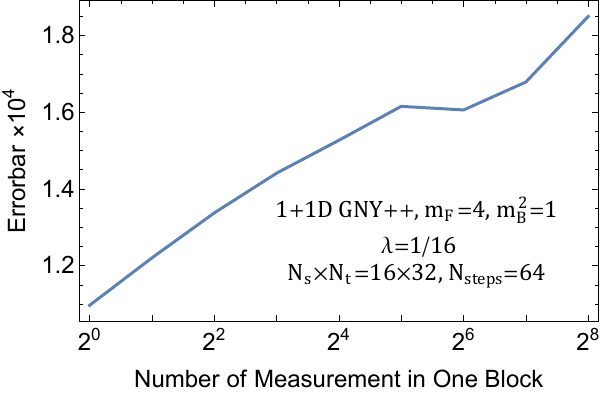}
\includegraphics[width=0.49\textwidth]{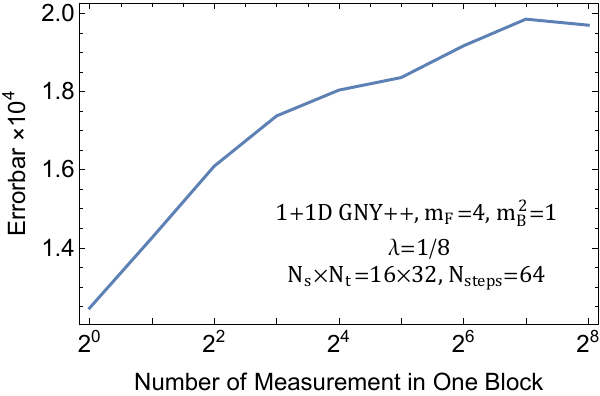}
\includegraphics[width=0.49\textwidth]{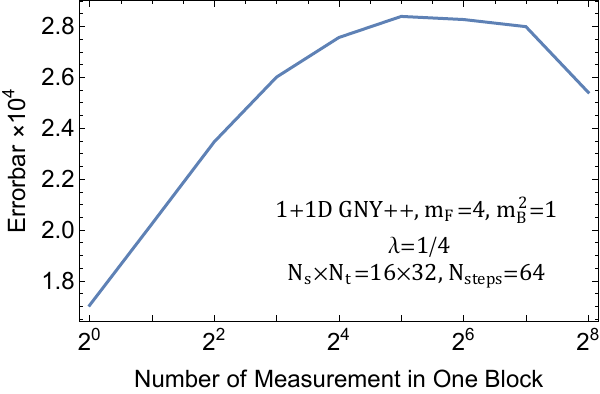}
\includegraphics[width=0.49\textwidth]{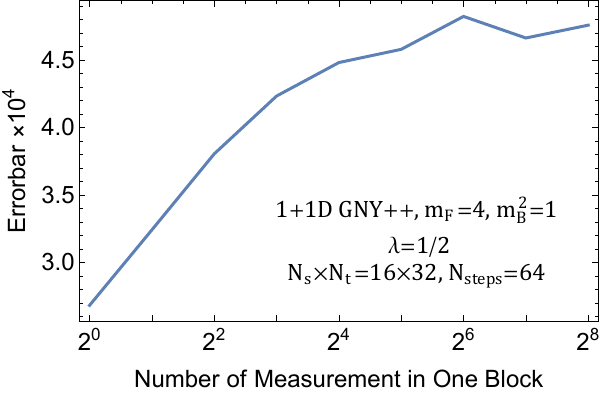}
\includegraphics[width=0.49\textwidth]{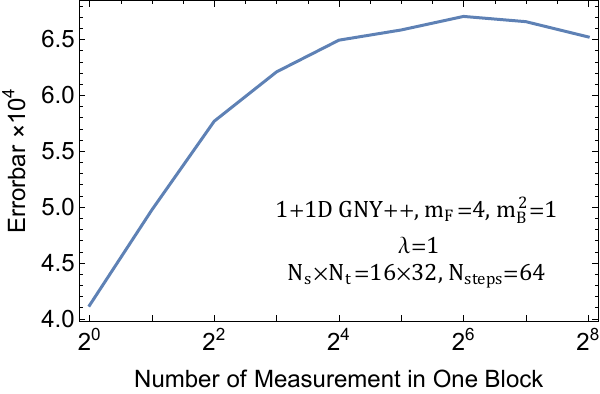}
\includegraphics[width=0.49\textwidth]{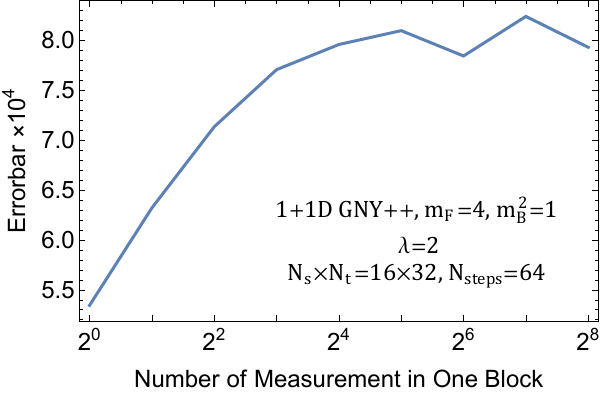}
\caption{The error bar of $\phi^2$ for $m_{\mathrm{F}}=4$, $m^2_{\mathrm{B}}=1$, and $\lambda=1/32, /16, 1/8, 1/4, 1/2, 1, 2$ with the lattice size $(N_s, N_t)=(16, 32)$ and the number of molecular dynamics steps $N_{\mathrm{steps}}=64$ in the 2D GNY model with the forward-forward finite-difference.
}
\label{EB2}
\end{figure}

\begin{figure}
\includegraphics[width=0.49\textwidth]{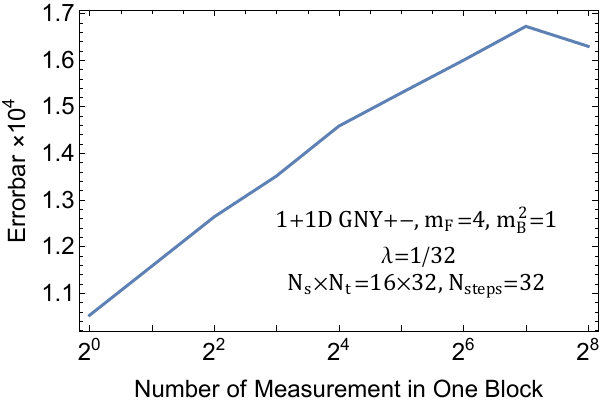}
\includegraphics[width=0.49\textwidth]{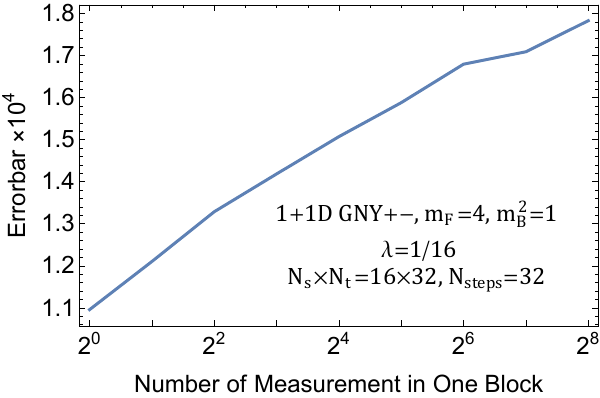}
\includegraphics[width=0.49\textwidth]{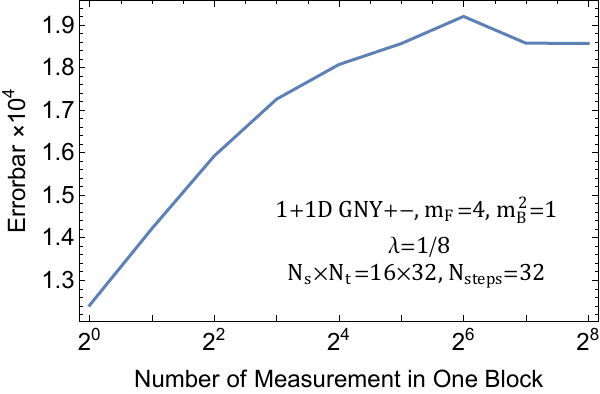}
\includegraphics[width=0.49\textwidth]{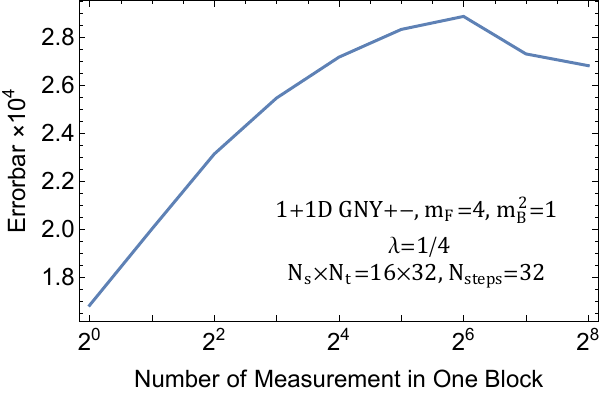}
\includegraphics[width=0.49\textwidth]{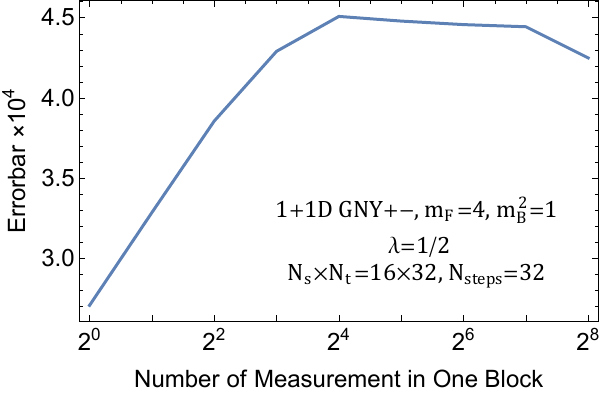}
\includegraphics[width=0.49\textwidth]{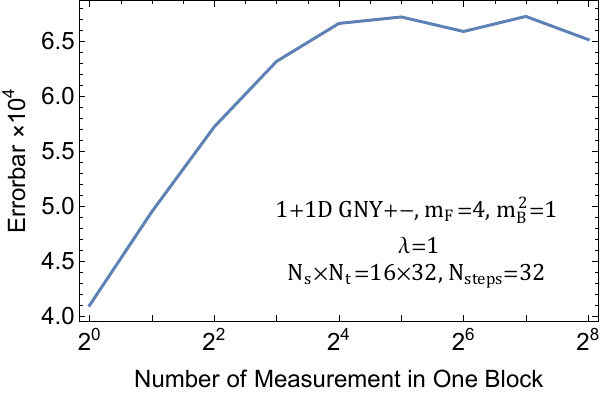}
\includegraphics[width=0.49\textwidth]{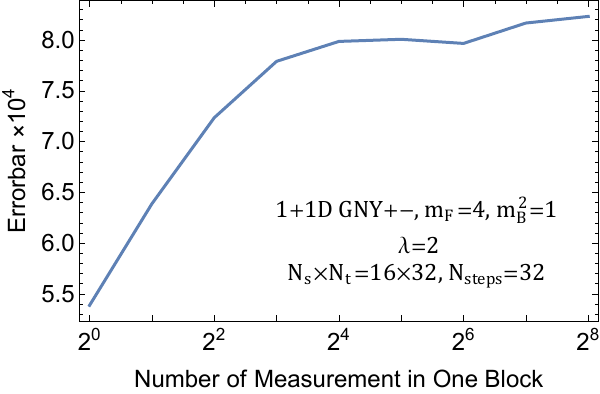}
\caption{The error bar of $\phi^2$ for $m_{\mathrm{F}}=4$, $m^2_{\mathrm{B}}=1$, and $\lambda=1/32, /16, 1/8, 1/4, 1/2, 1, 2$ with the lattice size $(N_s, N_t)=(16, 32)$ and the number of molecular dynamics steps $N_{\mathrm{steps}}=32$ in the 2D GNY model with the forward-backward finite-difference.
}
\label{EB3}
\end{figure}

\begin{figure}
\includegraphics[width=0.49\textwidth]{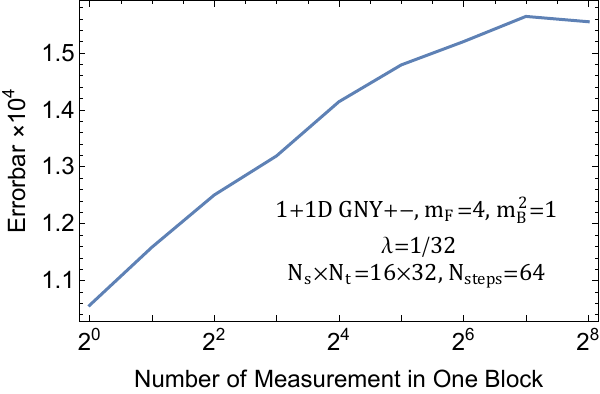}
\includegraphics[width=0.49\textwidth]{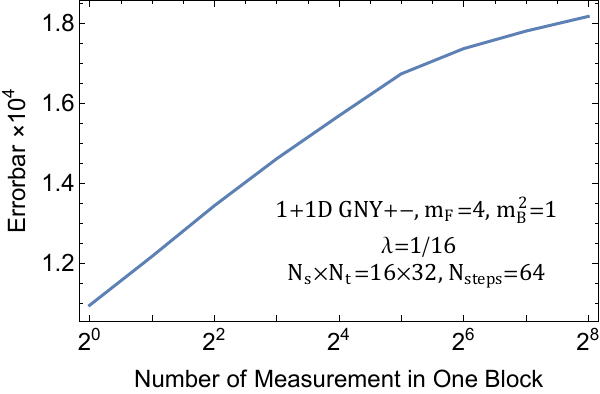}
\includegraphics[width=0.49\textwidth]{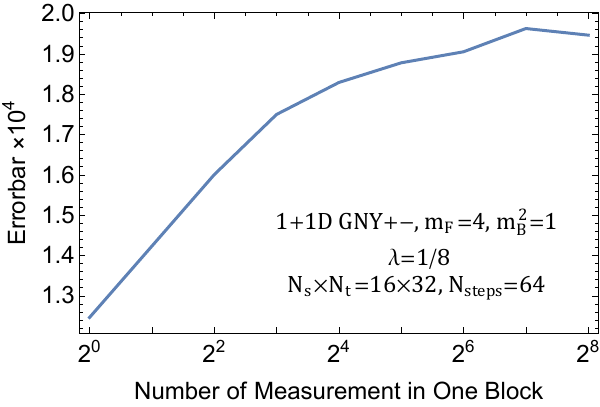}
\includegraphics[width=0.49\textwidth]{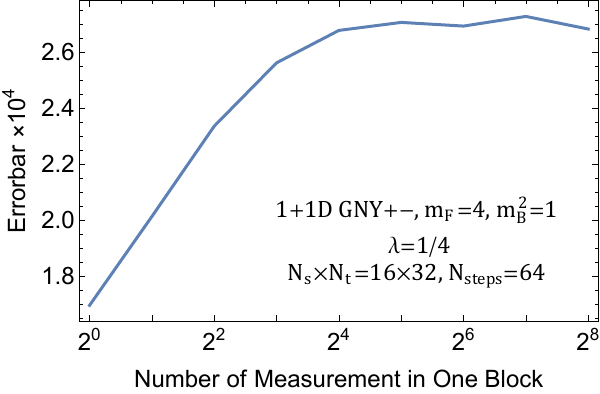}
\includegraphics[width=0.49\textwidth]{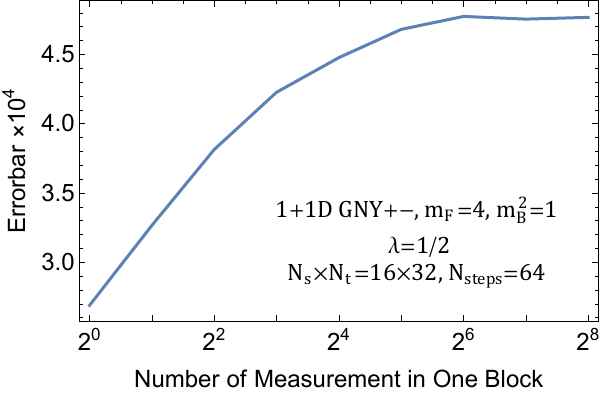}
\includegraphics[width=0.49\textwidth]{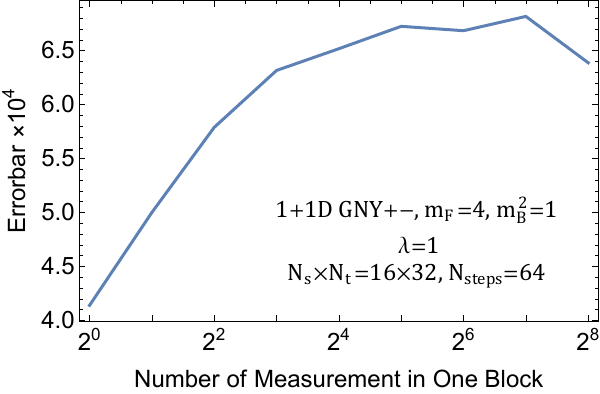}
\includegraphics[width=0.49\textwidth]{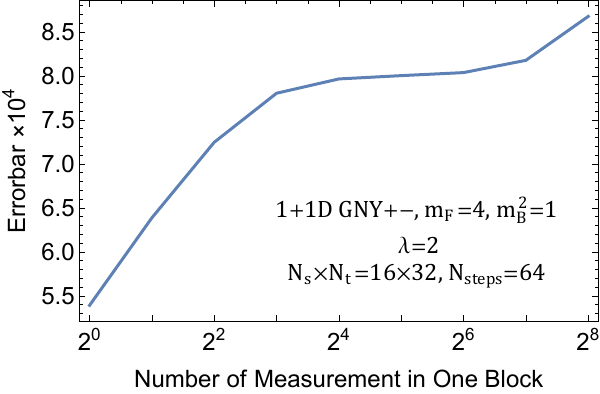}
\caption{The error bar of $\phi^2$ for $m_{\mathrm{F}}=4$, $m^2_{\mathrm{B}}=1$, and $\lambda=1/32, /16, 1/8, 1/4, 1/2, 1, 2$ with the lattice size $(N_s, N_t)=(16, 32)$ and the number of molecular dynamics steps $N_{\mathrm{steps}}=64$ in the 2D GNY model with the forward-backward finite-difference.
}
\label{EB4}
\end{figure}

\noindent 
We show the error bar of $\phi^2$ without the quenched averaging in Figs. \ref{EB1}, \ref{EB2}, \ref{EB3}, and \ref{EB4}.
The analysis of the error bar justifies the small auto-correlation by the measure interval, 32 sweeps. 

\end{document}